\documentclass[conference]{IEEEtran}
\IEEEoverridecommandlockouts
\usepackage{amsmath, amsthm, amssymb}
\usepackage{verbatim}
\usepackage{graphicx}
\usepackage{cite}

\usepackage[usenames]{color}
\definecolor{plum}  {rgb}{.4,0,.4}
\definecolor{BrickRed} {rgb}{0.6,0,0}
\usepackage[plainpages=false,pdfpagelabels,colorlinks=true,linkcolor=BrickRed,citecolor=plum]{hyperref}

\usepackage{hypcap}

\newtheorem{theorem}{Theorem}
\newtheorem{lemma}{Lemma}

\newtheorem{corollary}{Corollary}
\newtheorem{example}{Example}

\newcommand{\R}{\mathbb{R}}
\newcommand{\E}{\mathbb{E}}

\newcommand{\I}{\mathbf{1}}

\def\cL{{\mathcal L}}

\def\sW{{\mathsf W}}
\def\sX{{\mathsf X}}
\def\sY{{\mathsf Y}}

\def\PP{{\mathbb P}}

\def\deq{{\triangleq}}
\def\wh#1{{\widehat{#1}}}

\def\eps{\varepsilon}
\begin{document}

\sloppy

\title{Converses for Distributed Estimation\\
via Strong Data Processing Inequalities}

\author{Aolin Xu and Maxim Raginsky
\thanks{The authors are with the Department of Electrical and Computer Engineering and the Coordinated Science Laboratory, University of Illinois, Urbana, IL 61801, USA. E-mails: \{aolinxu2,maxim\}@illinois.edu.}
\thanks{Research supported in part by the NSF under award no.\ CCF-1017564, by CAREER award no.\ CCF-1254041, and by the Center for
Science of Information (CSoI), an NSF Science and Technology Center, under grant agreement CCF-0939370.}}
\maketitle

\begin{abstract}
We consider the problem of distributed estimation, where local processors observe independent samples conditioned on a common random parameter of interest, map the observations to a finite number of bits, and send these bits to a remote estimator over independent noisy channels. We derive converse results for this problem, such as lower bounds on Bayes risk.
The main technical tools include a lower bound on the Bayes risk via mutual information and small ball probability, as well as strong data processing inequalities for the relative entropy. Our results can recover and improve some existing results on distributed estimation with noiseless channels, and also capture the effect of noisy channels on the estimation performance.

\end{abstract}

\section{Introduction}
The problem of distributed estimation arises when the estimator does not have direct access to the samples generated according to the parameter of interest, but only to the data received from local processors that observe the samples.
In this work, we consider a general model of distributed estimation, where each of the $m$ processors observes $n$ independent samples drawn conditionally on a common $d$-dimensional parameter, generates a $b$-bit quantized message, and sends it to a remote estimator 
with $T$ uses of an independent noisy channel. We derive lower bounds on the Bayes risk and  on the minimum $b$ or $T$ needed to achieve a certain Bayes risk.
Fundamental limits of similar problems have been studied recently by Duchi et al.~\cite{Duchi_dist} and Shamir \cite{Shamir_dist14} with the assumption of noiseless channels (cf.~also earlier work by Gallager \cite{Gallager88} and by Han and Amari \cite{Han_inf98}).

To some extent, the parameter to be estimated in the problem under consideration can be viewed as a message to be sent in a transmission system, and the samples to be processed and quantized can be viewed as the input data to a compression system. 
However, a few important features make the problem distinct from data compression and transmission.
First, the dimension of the parameter may be fixed and not grow with the number of channel uses. 
Second, due to communication and computation constraints, the number of bits in the quantized message may not grow with the sample size. For example, as pointed out in \cite{Han_inf98}, the samples can be compressed at asymptotically zero rate, which makes it impossible to reconstruct the samples, yet still suffices to reliably estimate the parameter.
The number of bits in the quantized message may not grow with the number of channel uses either.
Due to these features, the conventional coding theorems in information theory cannot be applied here, but we can still use information-theoretic techniques to derive fundamental limits for the general problem of distributed estimation.

One of the major tools we use is a lower bound on the Bayes risk in terms of mutual information and small ball probability, which we derive using techniques introduced in our earlier work \cite{ISIT14_dist_comp}.
Another major tool is the strong data processing inequality (SDPI) for relative entropy \cite{Cohen_SDPI,VA_SDPI,MR_SDPI}, which lets us quantify the contraction of mutual information caused by communication constraints. 

The general results we obtain are non-asymptotic in $d$, $n$, $b$, $T$ and $m$, and can be used to derive asymptotic results. Examples are given for estimating both discrete and continuous parameters, where the converses closely match achievable performance. Moreover, our results can be naturally applied to minimax lower bounds, since the latter are always lower-bounded by the Bayes risk. We start with the single-processor setting, and then generalize the results to the multi-processor setting.
We are able to recover and improve some existing results on distributed estimation with noiseless channels \cite{Duchi_dist} as special cases, while our general results can capture the effect of noisy channels on the estimation performance.

\section{Main tools}

\noindent{\em A lower bound on Bayes risk.} In the standard Bayesian estimation framework, $\mathcal P = \{P_{X|W=w}: w\in\sW\}$ is a family of distributions on an observation space $\sX$, and the parameter space $\sW$ is endowed with a prior $P_W$.
 We estimate $W$ from $X \sim P_{X|W}$ as $\wh W = \psi(X)$, via an estimator $\psi$.
Given a distortion function $\ell:\sW\times\sW \rightarrow \R^+$, define the Bayes risk
\begin{align*}
R_{\rm B} = \inf_\psi \E\big[\ell(W,\wh W)\big].
\end{align*}
For a given $\psi$, the excess distortion probability $\PP(\ell(W,\wh W) > \rho)$ can be lower bounded in terms of the mutual information $I(W;\wh W)$ and the so-called \textit{small ball probability} of $W$ with respect to distortion function $\ell$ \cite{ISIT14_dist_comp}, defined as 
\begin{align*}
\cL(W,\rho) = \sup_{w\in\sW}\PP\big(\ell(W,w) \le \rho\big) .
\end{align*}
This quantity measures the ``spread'' of the prior distribution $P_W$. The lower bound on $\PP(\ell(W,\wh W) > \rho)$ can be conveniently converted to a lower bound on $\E[\ell(W,\wh W)]$ through Markov's inequality. Using the techniques from our earlier work \cite{ISIT14_dist_comp}, we obtain the following lower bound on the Bayes risk (see Appendix~\ref{appd:RBayes_lb} for the proof): 
\begin{theorem}\label{th:RBayes_lb}
In the above Bayesian estimation framework, 
\begin{align*}
R_{\rm B} &\ge \sup_{\rho>0} \rho \left(1 - \frac{I(W;X) + \log2}{\log (1/\cL(W,\rho))} \right) .
\end{align*}
\end{theorem}
Similar methods to derive Bayes risk lower bounds have been recently proposed by Chen et al. \cite{Chen_Bayes}, where they obtained lower bounds in terms of general $f$-informativities \cite{Csiszar72} and a quantity essentially the same as the small ball probability.
Theorem~\ref{th:RBayes_lb} reveals two sources of the intrinsic difficulty of estimating $W$: the amount of information about $W$ contained in the observation $X$, captured by $I(W;X)$, and the spread of the prior distribution $P_W$, captured by $\cL(W,\cdot)$. When an estimator does not have direct access to $X$ but only through one or more local processors, the mutual information between $W$ and the estimator's indirect observations will be a contraction of $I(W;X)$. The contraction is caused by the communication constraints between the local processors and the estimator, such as storage limitations of intermediate results, limited transmission blocklength, channel noise, etc. 

\medskip

\noindent{\em Contraction of mutual information via SDPI.} We quantify the contraction of mutual information using \textit{strong data processing inequalities} for the relative entropy (see \cite{MR_SDPI} and references therein).
Given a stochastic kernel (channel) $K$ with input alphabet $\sX$ and output alphabet $\sY$, and a reference input distribution $\mu$ on $\sX$, we say that $K$ satisfies an SDPI at $\mu$ with constant $c\in[0,1)$ if $D(\nu K \| \mu K) \le c D(\nu \| \mu)$ for any other input distribution $\nu$ on $\sX$. Here, $\mu K$ denotes the marginal distribution of the channel output when the input has distribution $\mu$. The tightest such constants,
\begin{align*}
\eta(\mu,K) \deq \sup_{\nu:\nu\neq\mu}\frac{D(\nu K \| \mu K)}{D(\nu \| \mu)},\,\,
\eta(K) \deq \sup_{\mu}\eta(\mu,K),
\end{align*}
are also the maximum contraction ratios of mutual information in a Markov chain \cite{VA_SDPI}:
for a Markov chain $W - X - Y$, 
\begin{align}\label{eq:MI_SDPI_Px}
\sup_{P_{W|X}} \frac{I(W;Y)}{I(W;X)} = \eta(P_X,P_{Y|X}) 
\end{align}
if the joint distribution $P_{X,Y}$ is fixed, and
\begin{align}\label{eq:MI_SDPI}
\sup_{P_{W,X}} \frac{I(W;Y)}{I(W;X)} = \eta(P_{Y|X}) 
\end{align}
if only the channel $P_{Y|X}$ is fixed. It is generally hard to precisely compute the SDPI constant for an arbitrary pair of $\mu$ and $K$, except for some special cases. One such case is that for binary symmetric channel, $\eta({\rm Bern}(\frac{1}{2}),{\rm BSC}(\eps)) = \eta({\rm BSC}(\eps)) = (1-2\eps)^2$ \cite{Ahlswede_Gacs_hypercont}. Various upper bounds on SDPI constants have been proposed (see \cite{MR_SDPI} and references therein). We will need one such bound \cite{Cohen_SDPI}:

\begin{lemma}\label{lm:Dobru_coef}
Define the {\em Dobrushin contraction coefficient} of a channel $P_{X|W}$ by $\vartheta(P_{X|W}) = \max_{w,w'}\|P_{X|W=w}-P_{X|W=w'}\|_{\rm TV}$.
Then $\eta(P_{X|W}) \le \vartheta(P_{X|W})$.
\end{lemma}
\noindent For product input distributions and product channels, the SDPI constant tensorizes \cite{VA_SDPI} (see \cite{MR_SDPI} for a more general result for other $f$-divergences):
\begin{lemma}\label{lm:SDPI_tensor}
For distributions $\mu_1,\ldots,\mu_n$ on $\sX$ and channels $K_1,
\ldots,K_n$ with input alphabet $\sX$,
\begin{align*}
\eta( \mu_1 \otimes \ldots \otimes \mu_n , K_1 \otimes \ldots \otimes K_n) = \max_{1\le i\le n} \eta(\mu_i, K_i) .
\end{align*}
\end{lemma}
\noindent Finally, motivated by Evans and Schulman \cite{Eva_Sch99} and by Polyanskiy and Wu \cite{PolWu_ES}, the following lemma characterizes the SDPI constant for multiple uses of a channel (see Appendix~\ref{appd:noisy_ch_SDPI} for the proof):
\begin{lemma}\label{lm:noisy_ch_SDPI}
For a stochastic kernel $P_{V|U}$, consider the stationary and memoryless channel $P_{V^T|U^T} = P_{V|U}^T$. The SDPI constant of $P_{V^T|U^T}$ satisfies
\begin{align*}
\eta(P_{V^T|U^T}) \le 1 - (1-\eta(P_{V|U}))^T .
\end{align*}
\end{lemma}

\section{Results for a single processor}\label{sec:single}

Consider the following distributed estimation problem with a single processor, shown schematically in Fig.~\ref{fg:model_single_bT}:

\begin{figure}[h!]
\centering
  \includegraphics[scale = 0.95]{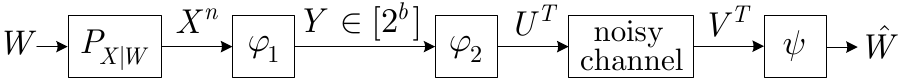}
  \caption{General model (single processor).}
  \label{fg:model_single_bT}
\end{figure}

\begin{itemize}
\item
$W = (W_1,\ldots,W_d)$ is a random parameter (discrete or continuous) with mutually independent coordinates.
\item
The $d\times n$ array of observations $X^n$ is generated conditionally on $W$ as follows: For each $j \in [d]$,
given $W_j=w_j$, the $n$ samples in the $j$th row of $X^n$, denoted by $X_j^n = (X_{j,1},\ldots,X_{j,n})$, are independently generated according to a given stochastic kernel $P_{X_j|W_j=w_j}$.
\item
The local processor observes $X^n$ and generates a $b$-bit message $Y=\varphi_1(X^n)$.
\item
The encoder maps $Y$ to a codeword $U^T = \varphi_2(Y)$ with blocklength $T$, and transmits $U^T$ over the noisy channel. The channel is memoryless, with stochastic kernel $P_{V|U}$. 
\item
The remote estimator $\psi$ estimates $W$ from the received codeword $V^T$, so that $\wh W = \psi(V^T)$.
\end{itemize}
The Bayes risk in this problem setup is defined as 
\begin{align*}
R_{\rm B} = \inf_{\varphi_1,\varphi_2,\psi} \E\big[\ell(W,\psi(V^T))\big].
\end{align*}
In order to apply Theorem~\ref{th:RBayes_lb}, we need an upper bound on the mutual information $I(W;V^T)$ which is independent of $\varphi_1$, $\varphi_2$, and $\psi$. All logarithms are binary, unless stated otherwise.
\begin{theorem}\label{th:gen_single}
For any choice of $\varphi_1$, $\varphi_2$, and $\psi$,
\begin{align*}
I(W;V^T) & \le \min\Big\{\left(H(X^n) \wedge b \right)\eta_T\max_{1\le j\le d}\eta(P_{X_j^n},P_{W_j|X_j^n}),\\
& \qquad \qquad I(W;X^n)\eta_T, CT \Big\}
\end{align*}
where $r \wedge s \deq \min\{r,s\}$, $\eta_T \deq 1-(1-\eta(P_{V|U}))^T$, and $C$ is the Shannon capacity of the channel $P_{V|U}$.
\end{theorem}
\begin{IEEEproof} Consider the Markov chain $W - X^n - Y - U^T - V^T$. From Lemma~\ref{lm:noisy_ch_SDPI}, Eq.~(\ref{eq:MI_SDPI}), and the ordinary DPI, we have
\begin{align}
I(W;V^T) \le I(W;U^T)\eta_T \le I(W;Y)\eta_T . \label{eq:MI_ub_WYWU}
\end{align}
On the one hand, 
\begin{align}\label{eq:I_WY_1}
I(W;Y) \le I(X^n;Y)\eta(P_{X^n},P_{W|X^n})
\end{align}
as a consequence of (\ref{eq:MI_SDPI_Px}).
Lemma~\ref{lm:SDPI_tensor} and the fact that $(W_1,X_1^n),\ldots,(W_d,X_d^n)$ are independent imply that
\begin{align*}
\eta(P_{X^n},P_{W|X^n}) = \max_{1\le j\le d}\eta(P_{X_j^n},P_{W_j|X_j^n}) 
\end{align*}
Finally, since $Y$ takes values in $[2^b]$,
\begin{align*}
I(X^n;Y)\le \min\{H(X^n),H(Y)\} \le \min\{H(X^n),b\}.
\end{align*}
Using these bounds in \eqref{eq:MI_ub_WYWU} and \eqref{eq:I_WY_1}, we get
\begin{align*}
	I(W;V^T) \le (H(X^n) \wedge b) \max_{1\le j\le d}\eta(P_{X_j^n},P_{W_j|X_j^n}) \eta_T.
\end{align*}
Alternatively, using $I(W;Y) \le I(W;X^n)$ in \eqref{eq:MI_ub_WYWU}, we get $I(W;V^T) \le I(W;X^n)\eta_T$. Lastly, because the noisy channel is memoryless, we have $I(W;V^T) \le I(U^T;V^T) \le CT$. We complete the proof by taking the minimum of the three resulting estimates to get the tightest bound on $I(W;V^T)$.
\end{IEEEproof}
Next we study a few examples of this problem setup to illustrate the effectiveness of using Theorem~\ref{th:RBayes_lb} and Theorem~\ref{th:gen_single} to derive converse results for the Bayes risk.

\medskip

\noindent{\bf Example 1: Transmitting a bit over a BSC.} Suppose $W$ is ${\rm Bern}(\frac{1}{2})$, $W=X^n=Y$, $P_{V|U}$ is ${\rm BSC}(\eps)$, so that $\eta(P_{V|U}) = (1-2\eps)^2$, and $\ell(w,\wh{w}) = {\bf 1}{\{w \neq \wh{w}\}}$.
\begin{corollary}\label{co:1bit_1sender}
The minimum blocklength $T^*$ to achieve $R_{\rm B} \le p$ satisfies
\begin{align*}
T^* \ge \frac{\log\frac{1}{h(p)}}{\log\frac{1}{4\eps\bar\eps}} 
\ge \frac{\log\frac{1}{p}-\log\log\frac{e}{p}}{\log\frac{1}{4\eps\bar\eps}} 
\sim \frac{\log\frac{1}{p}}{\log\frac{1}{4\eps\bar\eps}} \,\,\text{as $p \rightarrow 0$},
\end{align*}
where $h(\cdot)$ is the binary entropy function, and $\bar{\eps} \deq 1 -\eps$.
\end{corollary}
\begin{IEEEproof}
In this case, we can bypass Theorem~\ref{th:RBayes_lb} by using the bound $1-h(\PP(\wh W\neq W)) \le I(W;V^T)$.
Theorem~\ref{th:gen_single} gives $I(W;V^T) \le I(W;X^n)\eta_T \le 1-(4\eps\bar\eps)^T$. 
We obtain the lower bound using the fact that $h(p) \le p\log\frac{e}{p}$.
\end{IEEEproof}
The blocklength of a repetition code with error probability of at most $p$ gives an upper bound on $T^*$. By the Chernoff bound \cite{Gallager_ITbook}, a blocklength-$T$ repetition code can achieve $\PP(\wh W\neq W) \le 2^{-\frac{T}{2}\log\frac{1}{4\eps\bar\eps}}$. Thus
\begin{align*}. 
T^* \le {2\log\tfrac{1}{p}}\big/{\log\tfrac{1}{4\eps\bar\eps}} . 
\end{align*}
We see that the upper and lower bounds on $T^*$ only differ by a factor of $2$ as $p\rightarrow 0$, and have the same dependence on $\eps$.

\medskip

\noindent{\bf Example 2: Estimating a discrete parameter.} Consider the case where $W$ is uniformly distributed on $\{\pm 1\}^d$ and $X^n\in\{\pm 1\}^{d\times n}$. Given some fixed $\delta\in[0,1]$, $P_{X_j|W_j}(x_{j,k}|w_j) = ({1+x_{j,k}w_j\delta})/{2}$ for $j\in\{1,\ldots, d\}$ and $k\in\{1,\ldots,n\}$. In other words, $P_{X_{j}|W_j}$ is $\rm{BSC}(\frac{1-\delta}{2})$. It follows that $X_{j,k}$ is uniform on $\{\pm 1\}$, and $P_{W_j|X_{j,k}}$ is $\rm{BSC}(\frac{1-\delta}{2})$ as well. Channel $P_{V|U}$ is assumed to be arbitrary.
\begin{corollary}\label{co:W_disc}
In Example 2, for $n=1$,
\begin{align}
I(W;V^T) &\le \min\Big\{(d \wedge b)\delta^2\eta_T,\, CT\Big\}. \label{eq:W_disc_n=1}
\end{align}
For $n>1$, with $\beta \deq \frac{1-\delta}{1+\delta}$ and $\xi_n \deq \frac{1-\beta^n}{1+\beta^n}$,
\begin{align*}
	I(W;V^T) \le \min\Big\{&\left(d(1+nh(\tfrac{1-\delta}{2})) \wedge b\right)\xi_n\eta_T,
	d\eta_T, CT\Big\} .
\end{align*}
\end{corollary}
\begin{IEEEproof}
For $n=1$, we have the exact SDPI constant $\eta(P_{X_j},P_{W_j|X_j}) = \delta^2$, due to the fact that $X_j$ is uniform on $\{\pm 1\}$ and $P_{W_j|X_{j}}$ is $\rm{BSC}(\frac{1-\delta}{2})$.

For $n>1$, by Lemma~\ref{lm:Dobru_coef},
\begin{align}\label{eq:eta_disc_TV}
\eta(P_{X^n_j},P_{W_j|X^n_j}) &\le \eta(P_{W_j|X^n_j}) \le \vartheta(P_{W_j|X^n_j}) ,
\end{align}
where the Dobrushin coefficient is computed in Appendix~\ref{appd:Dbrsh_W_X} to be $\vartheta(P_{W_j|X^n_j}) = ({1-\beta^n})/({1+\beta^n})$. We also have $I(W;X^n)\le d$, and
\begin{align*}
H(X^n) &= dH(X^n_1) \le dH(W_1,X^n_1) \\
&= d(H(W_1)+H(X^n_1|W_1)) = d(1+nh(\tfrac{1-\delta}{2})) .
\end{align*}
The results then follow from Theorem~\ref{th:gen_single}.
\end{IEEEproof}
Duchi et al. \cite{Duchi_dist} considered the same problem with $n=1$ and noiseless $P_{V|U}$. Their result (Lemma 3 in \cite{Duchi_dist}), proved in a much more complicated way, shows that
\begin{align}
I(W;Y) &\le \min\{d,b\}{32\delta^2}/{(1-\delta)^4} \label{eq:W_disc_Duchi}
\end{align}
where the contraction coefficient is less than $1$ only when $\delta<0.133$.
In contrast, the contraction coefficient in (\ref{eq:W_disc_n=1}) can never go greater than $1$, and it considerably improves the contraction coefficient in (\ref{eq:W_disc_Duchi}) over all $\delta\in[0,1]$, especially for large $\delta$, under the same noiseless channel assumption.
Combined with Theorem~\ref{th:RBayes_lb}, Corollary~\ref{co:W_disc} can be applied to derive lower bounds on the minimax risk in estimating the mean of an arbitrary probability distribution on the cube $[-1,1]^d$. We discuss this application in Sec.~\ref{sec:multi}, in the multi-processor setting.

Using Corollary~\ref{co:W_disc}, we can obtain lower bounds on the bit error probability for estimating $W$ and on the number of bits to quantize the message $Y$.
\begin{corollary}
In Example 2, let $\ell(w,\wh{w}) = \tfrac{1}{d}\sum^d_{j=1} {\bf 1}{\{w_j \neq \wh{w}_j\}}$. Then, for $n=1$,
\begin{align*}
R_{\rm B} \ge h^{-1}\left(1-\frac{1}{d}\min \left\{ b\delta^2\eta_T,CT\right\}\right),
\end{align*}
provided $b$, $d$, and $T$ are such that the argument of $h^{-1}(\cdot)$ lies in $[0,1]$.
\end{corollary}
\begin{IEEEproof}
Let $d_2(\cdot\|\cdot)$ be the binary divergence function; then, choosing $\varphi_1,\varphi_2,\psi$ that attain $R_{\rm B}$, we have
\begin{align*}
&1-h(R_{\rm B}) = d_2(R_{\rm B}\| \tfrac{1}{2}) \le \frac{1}{d}\sum_{j=1}^d d_2(\PP(W_j \neq \wh W_j) \| \tfrac{1}{2}) \\
&\le \frac{1}{d}\sum_{j=1}^d I(W_j;\wh W_j) \le \frac{1}{d}I(W;\wh W)\le \frac{1}{d}\left(b\delta^2 \eta_T \wedge CT\right),
\end{align*}
where the first line uses the convexity of divergence, and the second line uses the data processing inequality for divergence, the fact that $W_j$'s are i.i.d., and Corollary~\ref{co:W_disc}. Applying $h^{-1}$ to both sides, we get the result.
\end{IEEEproof}

\begin{figure}[h!]
\centering
  \includegraphics[scale = 0.69]{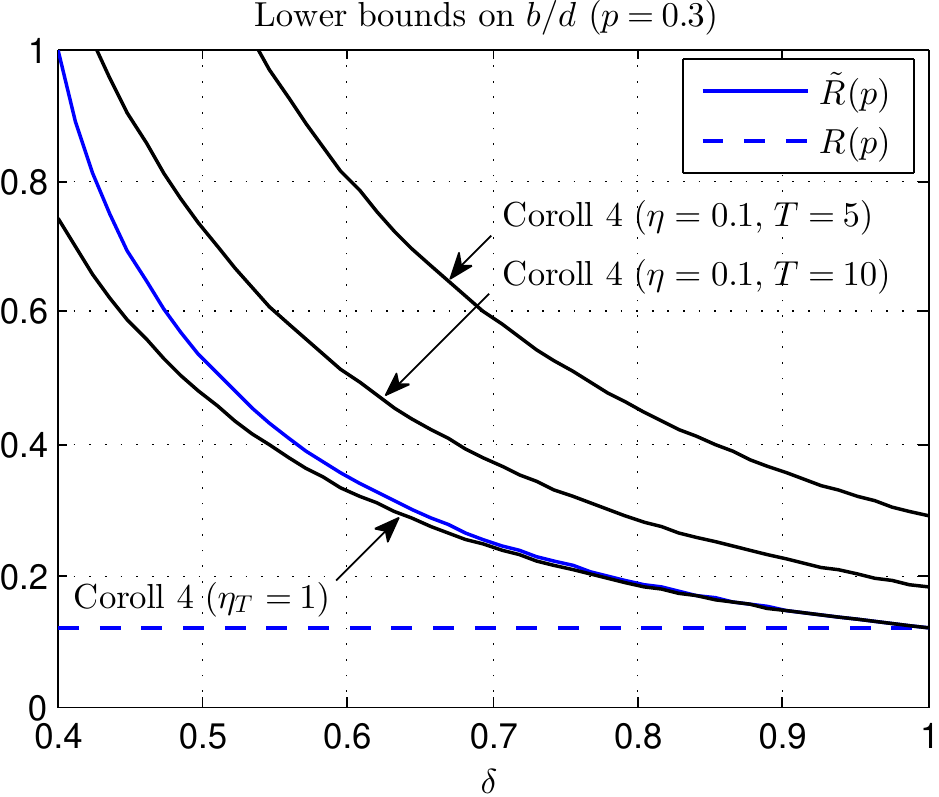}
  \caption{Comparison of lower bounds on $b/d$ ($p=0.3$).}
  \label{fg:noisy_lossy}
\end{figure}

\begin{corollary}
In Example 2, for $n=1$, to achieve $R_{\rm B} \le p$, it is necessary that
\begin{align*}
\frac{b}{d} \ge \frac{1-h(p)}{\delta^2 \eta_T} = \frac{1-h(p)}{\delta^2 \left(1-(1-\eta(P_{V|U}))^T\right)}.
\end{align*}
\end{corollary}
\noindent In Fig.~\ref{fg:noisy_lossy}, this lower bound is compared with the asymptotic compression ratio $\tilde R(p)= 1-h(\frac{2p+\delta-1}{2\delta})$, $0\le \frac{1-\delta}{2}\le p \le \frac{1}{2}$, of noisy lossy coding of an i.i.d.\ ${\rm Bern}(\frac{1}{2})$ source over ${\rm BSC}(\frac{1-\delta}{2})$, and also with the rate-distortion function $R(p)=1-h(p)$ of ${\rm Bern}(\frac{1}{2})$. 


\noindent {\bf Example 3: Estimating a continuous parameter.}
Consider the problem of estimating the bias of a Bernoulli random variable through a BSC. In this case, $W$ is assumed to be uniformly distributed on $[0,1]$, $P_{X|W=w}$ is ${\rm Bern}(w)$, and $P_{V|U}$ is ${\rm BSC}(\eps)$. We are interested in lower-bounding the Bayes risk with respect to the absolute loss $\ell(w,\wh{w}) = |w-\wh{w}|$.
\begin{corollary}\label{co:RB_cont} In Example~3, let $I^* \deq \sup_{\varphi_1,\varphi_2,\psi} I(W;V^T)$. Then the Bayes risk can be lower-bounded by
\begin{align}
R_{\rm B} \ge \frac{1}{16} 2^{-2I^*}  \label{eq:RB_cont_all_n}
\end{align}
for all values of $I^*$, and by
\begin{align}
R_{\rm B} \gtrsim \frac{1}{8I^*}2^{-I^*} \label{eq:RB_cont_asymp}
\end{align}
for $I^*\rightarrow \infty$. The notation $y \gtrsim g(x)$ means that there exists some function $f$ such that $y \ge f(x)$ for all $x$, and $\lim_{x\rightarrow\infty}{f(x)}/{g(x)} = 1$.
\end{corollary}
\begin{IEEEproof}
We have $\cL(W,\rho) = \sup_{w\in[0,1]}\PP\big(|W-w|<\rho\big) \le \min\{2\rho,1\}$.
From Theorem~\ref{th:RBayes_lb},
\begin{align}
R_{\rm B} \ge \sup_{0<\rho<\frac{1}{2}} \rho \left(1-\frac{I^*+\log 2}{\log (1/2\rho)}\right) \ge \frac{1}{2}\sup_{0<s<1} s 2^{-\frac{I^*+1}{1-s}}  \label{eq:MAE_lb_Bern_U}
\end{align}
where the last inequality is obtained by requiring $1-\frac{I^*+\log 2}{\log (1/2\rho)} >s$ for each $s\in(0,1)$. In (\ref{eq:MAE_lb_Bern_U}), taking $s=\frac{1}{2}$, we get (\ref{eq:RB_cont_all_n}), while
optimizing over $s$ and sending $I^*\rightarrow\infty$, we get (\ref{eq:RB_cont_asymp}).
\end{IEEEproof}
\begin{corollary}\label{co:I_cont}
In Example 3, for any choice of $\varphi_1,\varphi_2,\psi$,
\begin{align*}
I(W;V^T) &\le \min\Big\{b(1-2^{-n})\big(1-(4\eps\bar\eps)^T\big), \\
 & \big(\tfrac{1}{2}\log n + \gamma_n\big)\big(1-(4\eps\bar\eps)^T\big),
(1-h(\eps))T \Big\},
\end{align*}
where $\lim_{n\rightarrow\infty}\gamma_n = c$ with some absolute constant $c$.
\end{corollary}
\begin{IEEEproof}
From Lemma~\ref{lm:Dobru_coef},
\begin{align}\label{eq:Dbrsh_U_Bern}
\eta(P_{W|X^n}) \le \vartheta(P_{W|X^n}) = 1-2^{-n} .
\end{align}
The proof of (\ref{eq:Dbrsh_U_Bern}) is in Appendix~\ref{appd:Dbrsh_U_Bern}. Moreover (see, e.g., \cite{Clarke_Barron})
\begin{align*}
I(W;X^n) = \frac{1}{2}\log n + \gamma_n.
\end{align*}
With these facts, the result follows from Theorem~\ref{th:gen_single}.
\end{IEEEproof}
Now we apply Corollaries~\ref{co:RB_cont} and \ref{co:I_cont} to two specific cases:

\smallskip

\noindent \emph{Case 1}: $\eps=0$, $T\ge b$. We have
\begin{align*}
R_{\rm B} &\gtrsim \frac{1}{8(1-2^{-n})b}2^{-(1-2^{-n})b} 
\sim \frac{1}{4\sqrt{n}\log n}
\end{align*}
for $b=\tfrac{1}{2}\log n$, and $n\rightarrow \infty$.
To obtain an upper bound on $R_{\rm B}$, consider the scheme where the local processor quantizes the sample mean
$\bar{X}^n = n^{-1}\sum^n_{j=1}X_j$ into $\tilde W$ using a uniform $b$-bit quantization of $[0,1]$, and the remote estimator sets $\wh W = \tilde W$. By the triangle inequality,
\begin{align*}
\E|W-\wh W| \le \E|W-\bar X^n| + \E|\bar X^n - \tilde W|
\le \frac{1}{\sqrt{6n}}+2^{-b}.
\end{align*}
Thus, for $b=\tfrac{1}{2}\log n$, $R_{\rm B} \le {1.41}/{\sqrt{n}}$, which only differs from the lower bound by a logarithmic factor as $n\rightarrow \infty$.

\smallskip

\noindent \textit{Case 2}: $\eps>0$, $b\ge\log(n+1)$. We have
\begin{align*}
R_{\rm B} 
&\gtrsim \max\Big\{\frac{c_1}{\eta_T \sqrt{n^{\eta_T}}\log n },\frac{2^{-(1-h(\eps))T}}{8(1-h(\eps))T}\Big\} \\
&\ge \frac{\alpha c_1}{\eta_T \sqrt{n^{\eta_T}}\log n } + \frac{\bar\alpha 2^{-(1-h(\eps))T}}{8(1-h(\eps))T},\,\text{$\forall \alpha\in[0,1]$},
\end{align*}
where $\eta_T = 1-(4\eps\bar\eps)^T$, and $c_1$ is an absolute constant.
Consider the scheme where the local processor computes the sample sum $S^n$, which is uniformly distributed on $\{0,\ldots,n\}$, represents it with $\log(n+1)$ bits, and transmits these bits over the channel using a blocklength-$T$ code. The remote estimator decodes $S^n$ as $\wh S^n$, and sets $\wh W = \wh S^n/n$. Then
\begin{align*}
\E|W-&\wh W| \le \E|W- S^n/n| + \E| S^n/n - \wh S^n/n| \\ 
&\le \frac{1}{\sqrt{6n}}+ \PP(S^n \neq \wh S^n) 
\le \frac{1}{\sqrt{6n}}+2^{-E_r\big(\frac{\log(n+1)}{T}\big)T},
\end{align*}
where $E_r(\cdot)$ is the random coding error exponent of ${\rm BSC}(\eps)$:
\begin{align*}
E_r\big(\tfrac{1}{T}\log(n+1)\big) = 1 - \log(1+\sqrt{4\eps\bar\eps}) - \tfrac{1}{T}\log(n+1)
\end{align*}
when $\frac{1}{T}\log(n+1) \le 1-h\big(\frac{\sqrt{\eps}}{\sqrt{\eps}+\sqrt{\bar\eps}}\big)$ \cite[p.~146]{Gallager_ITbook}.
Note that 
\begin{align*}
1\le \frac{1-h(\eps)}{1 - \log(1+\sqrt{4\eps\bar\eps})} \le 2,\,\,\text{$\forall \eps\in(0,\tfrac{1}{2})$},
\end{align*}
which implies that the error exponent in the lower bound can closely match that in the upper bound at low transmission rate.

\section{A result for multiple processors}\label{sec:multi}

We now consider a set-up with $m$ local processors. Each processor observes an independent set of samples generated from a common random parameter $W$, and communicates with the remote estimator over an independent noisy channel. For notational simplicity, we assume that each processor applies the same local transformation to its samples, and that the channels between each processor and the remote estimator have the same transition probabilities. The results can be straightforwardly generalized to the case where the local encoders and the channels are different across the processors. 
\begin{theorem}\label{th:gen_multi}
In the multi-processor setup described above,
\begin{align*}
I(W;V^{m\times T})& \le m\min\Big\{I(W;X^n)\eta_T, \,CT, \\
&\big(H(X^n) \wedge b\big)\max_{1\le j\le d}\eta(P_{X_j^n},P_{W_j|X_j^n})\eta_T \Big\}.
\end{align*}
\end{theorem}
\begin{IEEEproof} Due to the independence assumption, the codewords $V^{m\times T}\deq (V^T_{(1)},\ldots,V^T_{(m)})$ received by the remote estimator from the processors $\{1,\ldots,m\}$ are conditionally independent given $W$. This implies that $I(W;V^{m\times T})\le \sum_{i=1}^m I(W;V^T_{(i)})$
(see, e.g., \cite[Lemma~4]{Duchi_dist}). Using Theorem~\ref{th:gen_single} to upper-bound each term, we obtain the result of Theorem~\ref{th:gen_multi}.
\end{IEEEproof}

Using Theorem~\ref{th:gen_multi} with Theorem~\ref{th:RBayes_lb} and Corollary~\ref{co:W_disc}, we can obtain a lower bound on the minimax risk for estimating the mean of an unknown distribution $P$ on $\sX = [-1,1]^d$, where each processor $i \in \{1,\ldots,m\}$ only observes a single independent sample $X_{(i)} \sim P$. Let $\mathcal P$ denote family of all probability distributions on $[-1,1]^d$. For $P\in\mathcal P$, the parameter of $P$ in this example is formally defined as $\theta(P) = \E_P[X]$. The minimax risk is defined as
\begin{align*}
R_{\rm M} = \inf_{\varphi_1^m,\varphi_2^m,\psi}\sup_{P\in \mathcal P} \E_P[\|\theta(P)- \psi(V^{m\times T})\|^2],
\end{align*}
where $\psi$ is an estimator of $\theta \in [-1,1]^d$.
\begin{corollary}\label{co:RM_lb_dist}
For the above minimax estimation problem,
\begin{align*}
R_{\rm M} \ge \frac{d}{6}\min\Big\{1,\frac{d}{24\eta_T m (d \wedge b)}\Big\}, \,\, \text{for $d\ge 12$}.
\end{align*}
\end{corollary}
\begin{IEEEproof}
At a high level, the proof strategy follows that in \cite{Duchi_dist}. But we use Theorem~\ref{th:RBayes_lb} instead of their distance-based Fano inequality, and we improve their mutual information upper bound by Corollary~\ref{co:W_disc}, which also captures the influence of noisy channels between the processors and the remote estimator. Let $W$, $\delta$, and $P_{X_j|W_j}$ be defined as in Example 2 with $n=1$. Given $W=w$, each processor observes a sample $X$ with its coordinates drawn according to $P_{X_j|W_j=w_j}$. Hence $P_{X|W=w}\in\mathcal P$ for all $w\in\{\pm 1\}^d$. 
Let $\theta_w \deq \theta(P_{X|W=w}) = \delta w$, so that $\|\theta_w-\theta_{w'}\|^2 = 4\delta^2\ell_{\rm H}(w,w')$, where $\ell_{\rm H}$ denotes the Hamming distance. 
Therefore,
\begin{align}\label{eq:RM_RB_dist}
R_{\rm M} &\ge 4\delta^2\inf_{\varphi^m_1,\varphi^m_2}\inf_{\psi}\E[\ell_{\rm H}(W,\wh W)],
\end{align}
where the second infimum is now over all remote estimators of $W \in \{-1,+1\}^d$. Let $\rho$ be a nonnegative integer. Then ${\cL}(W,\rho) = \sum_{\tau=0}^\rho{d \choose \tau} /2^d$, and $\log(1/{\cL}(W,\rho))\ge d/6$ for $\rho\le d/6$ and $d\ge 12$.
Thus from Theorem~\ref{th:RBayes_lb},
\begin{align*}
\inf_{\varphi^m_1,\varphi^m_2}\inf_{\psi}\E[\ell_{\rm H}(W,\wh W)] &\ge \rho\Big(1-\frac{I(W;V^{m\times T})+\log 2}{\log (1/{\cL}(W,\rho))}\Big) \\
&\ge \frac{d}{6}\Big(1-\frac{I(W;V^{m\times T})+\log 2}{d/6}\Big) .
\end{align*}
From Corollary~\ref{co:W_disc} and Theorem~\ref{th:gen_multi}, we have $I(W;V^{m\times T}) \le m \delta^2 \eta_T \min\{d,b\}$. Thus from (\ref{eq:RM_RB_dist}),
\begin{align*}
R_{\rm M} &\ge \frac{2d\delta^2}{3}\Big(1-\frac{m\delta^2\eta_T (d \wedge b)+\log 2}{d/6}\Big) .
\end{align*}
With $\delta^2 = \min\{1,{d}/({24 m\eta_T (d \wedge b)})\}$, the quantity in the parentheses is $\ge 1/4$, and we obtain the desired result.
\end{IEEEproof}
In the noiseless channel case, Corollary~\ref{co:RM_lb_dist} recovers and improves the lower bound in Proposition~2 of \cite{Duchi_dist}, which can be achieved within a constant factor using a method described there. When the processors communicate to the estimator via noisy channels, the effect of the noise on the minimax risk is captured by $\eta_T$ in the denominator of the lower bound.

\section*{Acknowledgment} The authors would like to thank Y.~Polyanskiy and Y.~Wu for helpful discussions and for making Ref.~\cite{PolWu_ES} available.



\clearpage

\appendices

\renewcommand{\theequation}{\Alph{section}.\arabic{equation}}
\renewcommand{\thelemma}{\Alph{section}.\arabic{lemma}}
\setcounter{equation}{0}
\setcounter{lemma}{0}

\section{Proof of Theorem~\ref{th:RBayes_lb}}\label{appd:RBayes_lb}
For any estimator $\psi$, let $P$ be the joint distribution of $W$ and $\wh W$, and $Q$ be the product of the marginals of $P$. Define $p_\rho = P(\ell(W,\wh W) \le \rho)$ and $q_\rho = Q(\ell(W,\wh W) \le \rho)$ for an arbitrary $\rho>0$. Then, by data processing inequality of divergence,
\begin{align}
I(W;\wh W) &= D(P \| Q) 
\ge d_2(p_\rho \| q_\rho) 
\ge p_\rho\log\frac{1}{q_\rho} - h(p_\rho) \nonumber \\
&\ge p_\rho\log\frac{1}{\cL(W,\rho)} - \log 2 \label{eq:pf_lm1_dp}
\end{align}
where the last inequality follows from the fact that 
\begin{align*}
q_\rho &= \int_{\sW}\int_\sW \I\{\ell(w,\wh w) \le \rho\}P_W({\rm d}w)P_{\wh W}({\rm d}\wh w) \\
&\le \sup_{\wh w\in\sW} \E[ \I\{\ell(W,\wh w) \le \rho\} ]
= \cL(W,\rho) .
\end{align*}
Consequently,
\begin{align}
1-p_\rho \ge 1 - \frac{I(W;\wh W) + \log2}{\log (1/\cL(W,\rho))} . \label{eq:1-pt_lb}
\end{align}
From the fact that $\ell(W,\wh W) \ge \rho \I\{\ell(W,\wh W) > \rho\}$, we have
\begin{align*}
\E[\ell(W,\wh W)] 
\ge \rho P(\ell(W,\wh W) > \rho) 
= \rho(1-p_\rho) .
\end{align*}
Lower bounding $1-p_\rho$ with (\ref{eq:1-pt_lb}), we get
\begin{align*}
\E[\ell(W,\wh W)] &\ge \sup_{\rho>0} \rho \left(1 - \frac{I(W;\wh W) + \log2}{\log (1/\cL(W,\rho))} \right) .
\end{align*}
The proof is completed by taking the infimum over $\psi$, and using the fact that $I(W;\wh W)\le I(W;X)$.

\section{Proof of Lemma~\ref{lm:noisy_ch_SDPI}}\label{appd:noisy_ch_SDPI}
Let $Y$ be an arbitrary random variable such that $Y \rightarrow U^T \rightarrow V^T$ form a Markov chain. Suppose $\eta(P_{V|U}) = \eta$. It suffices to show that
\begin{align}
I(Y;V^T) \le \big(1-(1-\eta)^T\big)I(Y;U^T) .\label{eq:I_YU_YV}
\end{align}
From the chain rule,
\begin{align*}
I(Y;V^T) = I(Y;V^{T-1}) + I(Y;V_T|V^{T-1}) .
\end{align*}
Since $Y,V^{T-1} \rightarrow U_T \rightarrow V_T$ form a Markov chain, a conditional version of SDPI (Corollary 1 in \cite{Eva_Sch99}) gives
\begin{align*}
I(Y;V_T|V^{T-1}) \le \eta I(Y;U_T|V^{T-1}) .
\end{align*}
It follows that
\begin{align*}
I(Y;V^T) &\le I(Y;V^{T-1}) + \eta I(Y;U_T|V^{T-1}) \\
&= (1-\eta)I(Y;V^{T-1}) + \eta I(Y;V^{T-1},U_T) \\
&\le (1-\eta)I(Y;V^{T-1}) + \eta I(Y;U^T) ,
\end{align*}
where the last step follows from the ordinary data processing inequality and the fact that $Y \rightarrow U^{T-1} \rightarrow V^{T-1}$ form a Markov chain. Unrolling the above recursive upper bound on $I(Y;V^T)$ and noting that $I(Y;V_1) \le I(Y;U_1)\eta$, we get
\begin{align*}
I(Y;V^T) &\le (1-\eta)^{T-1}\eta I(Y;U_1) + \ldots + \\
&\quad\,\, (1-\eta)\eta I(Y;U^{T-1}) + \eta I(Y;U^T) \\
&\le \big((1-\eta)^{T-1}+ \ldots + (1-\eta)+1\big)\eta I(Y;U^T) \\
&= \big(1-(1-\eta)^T\big) I(Y;U^n),
\end{align*}
which proves (\ref{eq:I_YU_YV}) and hence Lemma~\ref{lm:noisy_ch_SDPI}.

\section{Proof of \eqref{eq:eta_disc_TV}} \label{appd:Dbrsh_W_X}
We have $P_{W_j}(w_j) = \frac{1}{2}$ for $w_j = \pm 1$, and 
\begin{align*}
P_{X_j^n|W_j}(x_j^n|w_j) = \left(\frac{1+w_j\delta}{2}\right)^s \left(\frac{1-w_j\delta}{2}\right)^{n-s},
\end{align*}
where $s$ is the number of $1$'s in $x_j^n$. Thus
\begin{align*}
P_{W_j|X_j^n}(w_j|x_j^n)&=  \frac{\left(\frac{1+w_j\delta}{2}\right)^s \left(\frac{1-w_j\delta}{2}\right)^{n-s}}{\left(\frac{1+\delta}{2}\right)^s \left(\frac{1-\delta}{2}\right)^{n-s} + \left(\frac{1-\delta}{2}\right)^s \left(\frac{1+\delta}{2}\right)^{n-s}} \\
&=  
\begin{cases}
\dfrac{1}{1+\beta^{2s-n}} ,& \text{if $w_j=1$} \\
\dfrac{1}{1+\beta^{-2s+n}}, & \text{if $w_j=-1$}
\end{cases} .
\end{align*}
This gives
\begin{align*}
&\|P_{W|X^n=x^n} - P_{W|X^n=\tilde x^n}\|_{\rm TV} = \\ 
& \frac{1}{2}\left( \Big| \dfrac{1}{1+\beta^{2s-n}} - \dfrac{1}{1+\beta^{2\tilde s-n}} \Big| + \Big| \dfrac{1}{1+\beta^{-2s+n}} - \dfrac{1}{1+\beta^{-2\tilde s+n}} \Big| \right) ,
\end{align*}
which is maximized by choosing $x_j^n$ and $\tilde x_j^n$ such that $s = 0$ and $\tilde s = n$. Hence
\begin{align*}
\vartheta(P_{W_j|X^n_j}) = \frac{1}{1+\beta^n} - \frac{1}{1+\beta^{-n}} 
= \frac{1-\beta^n}{1+\beta^n} .
\end{align*}

\section{Proof of (\ref{eq:Dbrsh_U_Bern})}\label{appd:Dbrsh_U_Bern}
We have $p_W(w) = 1$ for $w\in[0,1]$, and $P_{X^n|W}(x^n|w) = w^s(1-w)^{n-s}$, where $s$ is the number of $1$'s in $x_n$. Thus 
\begin{align*}
P_{X^n}(x^n) = \int_0^1 w^s(1-w)^{n-s}{\rm d}w = \frac{1}{(n+1){n \choose s}}
\end{align*}
and
\begin{align*}
P_{W|X^n}(w|x^n) =  w^s(1-w)^{n-s}(n+1){n \choose s} .
\end{align*}
This gives
\begin{align*}
\|&P_{W|X^n=x^n} - P_{W|X^n=\tilde x^n}\|_{\rm TV} = \\ 
&\frac{n+1}{2}\int_0^1 \Big| w^s(1-w)^{n-s}{n \choose s} - w^{\tilde s} (1-w)^{n-\tilde s}{n \choose \tilde s} \Big| {\rm d}w ,
\end{align*}
which is maximized by choosing $x^n$ and $\tilde x^n$ such that $s = 0$ and $\tilde s = n$. Hence
\begin{align*}
\vartheta(P_{W|X^n}) &= \frac{n+1}{2}\int_0^1 \big| (1-w)^{n} - w^{n} \big| {\rm d}w  
= 1-2^{-n}.
\end{align*}

\end{document}